\documentclass[final,5p,twocolumn]{elsarticle}

\usepackage{lineno,hyperref}

\usepackage{upgreek}
\usepackage{amssymb}
\usepackage{booktabs}

\newcommand{\atsupp}{Appendix}

\biboptions{sort&compress}

\makeatletter
\def\ps@pprintTitle{%
  \let\@oddhead\@empty
  \let\@evenhead\@empty
  \def\@oddfoot{\reset@font\hfil\thepage\hfil}
  \let\@evenfoot\@oddfoot
}
\makeatother

\begin{document}

\begin{frontmatter}

\title{Kinetically limited valence of colloidal particles\\with surface mobile {DNA} linkers}

\author{Pedro A. S\'anchez\corref{mycorrespondingauthor}}
\cortext[mycorrespondingauthor]{Corresponding author}
\ead{pedro.sanchez@univie.ac.at}
\address{Computational and Soft Matter Physics, Faculty of Physics, University of Vienna, 1090 Vienna, Austria}

\author{Alessio Caciagli}
\address{Former affiliated: Department of Physics, University of Cambridge, Cambridge, CB3 0HE, United Kingdom}

\author{Sofia S. Kantorovich}
\address{Computational and Soft Matter Physics, Faculty of Physics, University of Vienna, 1090 Vienna, Austria}
\address{Research Platform MMM Mathematics-Magnetism-Materials, Vienna, Austria}

\author{Erika Eiser}
\address{Department of Physics, Norwegian University of Science and Technology, 7491 Trondheim, Norway}
\address{Department of Physics, University of Cambridge, Cambridge, CB3 0HE, United Kingdom}

\begin{abstract}
We characterize the self-assembly of colloidal particles with surface mobile DNA linkers under kinetically limited valence conditions. For this, we put forward a computer simulation model that captures quantitatively the interplay between the main dynamic processes governing these systems and allows the simulation of the long time scales reached in experiments. The model is validated by direct comparison with available experimental results, showing an overall good agreement that includes measurements of the average effective valence and its probability distribution as a function of the density of DNA linkers on the particles surface. Finally, simulation results are used to evidence the opposite impact of particle density and characteristic DNA hybridization time on the effective valence.
\end{abstract}

\end{frontmatter}

\section{Introduction}

Current experimental techniques allow the synthesis of colloidal particles with finely tuned selectivity and directionality of their bonding interactions. This makes such particles outstanding potential building blocks for the rational design of material created by hierarchical self-assembly \cite{2015-cademartiri-nm, 2016-yan, 2017-ravaine-cocis, 2017-teixeira-cocis, 2017-bianchi-pccp}. One of the most interesting strategies to design colloidal particles with directed self-assembly properties is based on their surface functionalization with DNA strands of selected base sequences. The bonding interactions of such DNA-coated colloids (DNACCs) are therefore mediated by DNA hybridization of complementary strands, which act as crosslinkers \cite{2012-wang-nat, 2016-rogers-nrm, 2019-laramy-nrm, 2021-zhang-fp}. The thermo-reversibility and high specificity of DNA hybridization provides precise temperature control of the colloidal self-assembly and a very flexible mechanism to tune the resulting structures and rheological properties \cite{2007-nykypanchuk-lm, 2008-nykypanchuk-nat, 2010-jones-nm, 2011-macfarlane-sc, 2013-dimichele-pccp, 2021-stoev-pre}.

Most early works on directed self-assembly of DNACCs were focused on the crystallization of solid core particles functionalized with arrangements of DNA linkers anchored to fixed points of their surfaces, first using nanoparticles \cite{1996-mirkin-nat, 1996-alivisatos-nat} and later larger colloidal particles as solid cores \cite{2003-milam-lm, 2005-biancaniello-prl, 2005-rogers-lm, 2006-kim-lm}. In the latter case, crystallization turned out to be more difficult due to the relatively shorter range of the DNA crosslinking interaction with respect to the particle size. This leads to rather rigid interparticle bonds and, therefore, to a higher tendency to form kinetically trapped amorphous aggregates \cite{2010-dreyfus-pre, 2012-angioletti-uberti-nm, 2013-dimichele-pccp}. Thus, regular structures of large colloidal particles have to be obtained by controlling the balance of binding/unbinding processes of the DNA linkers, so that reconfigurations of the aggregates by means of `rolling motions' of the bonded particles (\textit{i.e.}, orbit-like relative movements) can take place \cite{2006-kim-lm, 2015-wang-nc}. This also requires a large and homogeneous enough surface coverage of linkers \cite{2015-wang-nc}. Nevertheless, dynamics of this crystallization process tend to be very slow \cite{2005-biancaniello-prl}, difficulting any large scale production.

The aforementioned limitations stimulated the exploration of novel strategies for the design of DNACCs. One of such approaches is the usage of soft materials as core particles. Currently there is a pletora of soft core DNACCs, including microemulsion droplets \cite{2009-dreyfus-prl, 2012-hadorn-pnas, 2013-feng-avm, 2013-feng-sm, 2016-joshi-sa, 2017-zhang-nc} vesicles \cite{2015-parolini-nc, 2016-hadorn-lm} and supramolecular DNA nanostructures \cite{2021-xiong-nl}. This opened up the possibility to create soft materials of high technological interest, such as thermo-reversible gels with tunable properties \cite{2018-caciagli-lm}. Another important recent development is the synthesis of DNACCs with on-surface mobile linkers, \textit{i.e.}, systems in which the anchoring points of the DNA chains can diffuse on the surface of the core particle. For solid cores this is achieved by coating them with a non-rigid interface, such as a lipid bilayer, to which the linkers are attached \cite{2013-meulen-jacs}. Many soft core particles, such as droplets, vesicles and micelles, have non-rigid interfaces that make linkers attached to them naturally mobile \cite{2013-feng-sm, 2017-jo-cocis, 2018-caciagli-lm}. Importantly, such mobility is not limited to unbinded linkers, as already established crosslinkers can still diffuse on the surface of the particles they bind. Surface diffusion tends to smear out inhomogeneities in the distribution of free linkers and allows rolling motions of the bonded particles without the need of any unbinding process, lighting up the conditions and time scales for crystallization. In addition, mobile linkers provide additional ways to control the valence of the particles \cite{2014-angioletti-uberti-prl, 2016-angioletti-uberti-pccp} and ease the design of sophisticated sequential self-assembly schemes \cite{2013-dimichele-nc}.

The advantages brought by the on-surface mobility of the linkers in DNACCs come at the cost of a more challenging theoretical characterization than the one required by their rigidly anchored counterparts. This, together with their earlier development, made the latter the main subject of theoretical studies on DNACCs, frequently addressed as a particular case of ligand-receptor assembling systems \cite{2006-paudry-pnas, 2008-lee-lm, 2011-leunissen-jcp, 2016-bachmann-sm, 2016-angioletti-uberti-pccp, 2019-jana-pre}. In general, such studies have been focused on the binding/unbinding dynamics of the linkers, as the main parameter determining the final assembled structure, and the strategies to avoid kinetically arrested configurations \cite{2009-dreyfus-prl, 2010-dreyfus-pre, 2012-angioletti-uberti-nm, 2012-mognetti-sm, 2013-wu-pre, 2014-jenkins-pnas, 2015-rogers-sc, 2015-theodorakis-cmp, 2016-bachmann-sm, 2016-angioletti-uberti-pccp, 2019-jana-pre, 2019-pretti-ms}. Except for some favorable limit cases---\textit{i.e.}, systems with linker lengths comparable to core sizes and very low number of linkers \cite{2006-starr-jpcm, 2010-hsu-prl}---most theortical models for DNACCs self-assembly rely on interparticle effective potentials that avoid any explicit linker representation \cite{2011-leunissen-jcp, 2013-dimichele-nc, 2019-pretti-ms}. Effective potentials are fitted to experimental data on the basis of configurational energy and combinatorial entropy considerations \cite{2012-varilly-jcp, 2013-mladek-sm, 2014-angioletti-uberti-prl, 2018-hu-prl}. However, in systems with surface mobile linkers, the interplay of the aforementioned processes with the diffusion of the linkers becomes determinant. The basic features of this interplay have been discussed in few pioneering works, mainly as a qualitative generalization of the findings obtained from specific experimental systems \cite{2013-feng-sm, 2016-joshi-sa}, whereas efforts towards a comprehensive theoretical characterization are still very scarce. Bachmann and co-workers presented an extension of the combinatorial thermodynamic approaches developed for systems with rigidly attached linkers to include two essential aspects of the kinetics of mobile linkers: first, the finite rate of linker binding processes and, second, the extinction of such processes once all available linkers become bonded. As a consequence of the competition between this model binding kinetics and the diffusion of the core particles, formation of clusters with very low average coordination number, or valence, was predicted \cite{2016-bachmann-sm}. However, as diverse experimental works have evidenced, this fits only to very dilute systems with very low surface densities of linkers, whereas more compact structures are obtained as particle and linker densities are increased \cite{2013-feng-sm, 2018-mcmullen-prl}. In one of such works, McMullen and co-workers used their experimental results to develop a two-fold phenomenological modelling approach that takes the experimental probability distributions for the coordination numbers as fitting parameters \cite{2018-mcmullen-prl}. A very recent work of this experimental group has addressed the self-assembly under thermodynamic equilibrium conditions---\textit{i.e.}, for temperatures around the hybridization threshold, $T_h$, corresponding to the binding/unbinding transition, and large concentrations of particles and linkers \cite{2021-mcmullen-pnas}. However, to our best knowledge, no theoretical study to date has been able to predict the structures observed in experiments under kinetically controlled self-assembly conditions  directly from the physical properties of the system \cite{2018-mcmullen-prl}. In this work we present a minimal computer simulation model for the stable self-assembly of micron-sized DNACCs with surface mobile linkers that, for the first time, takes into account all the relevant dynamic processes governing these systems and considers the time dependence of the effective interactions. Despite the model was originally aimed at the qualitative description of the interplay of such processes, it is able to provide quantitative predictions in reasonably good agreement with available experimental results.

\section{\label{sec:system}System qualitative description}

\begin{figure}[!t]
\centering
\includegraphics[width=8.7cm]{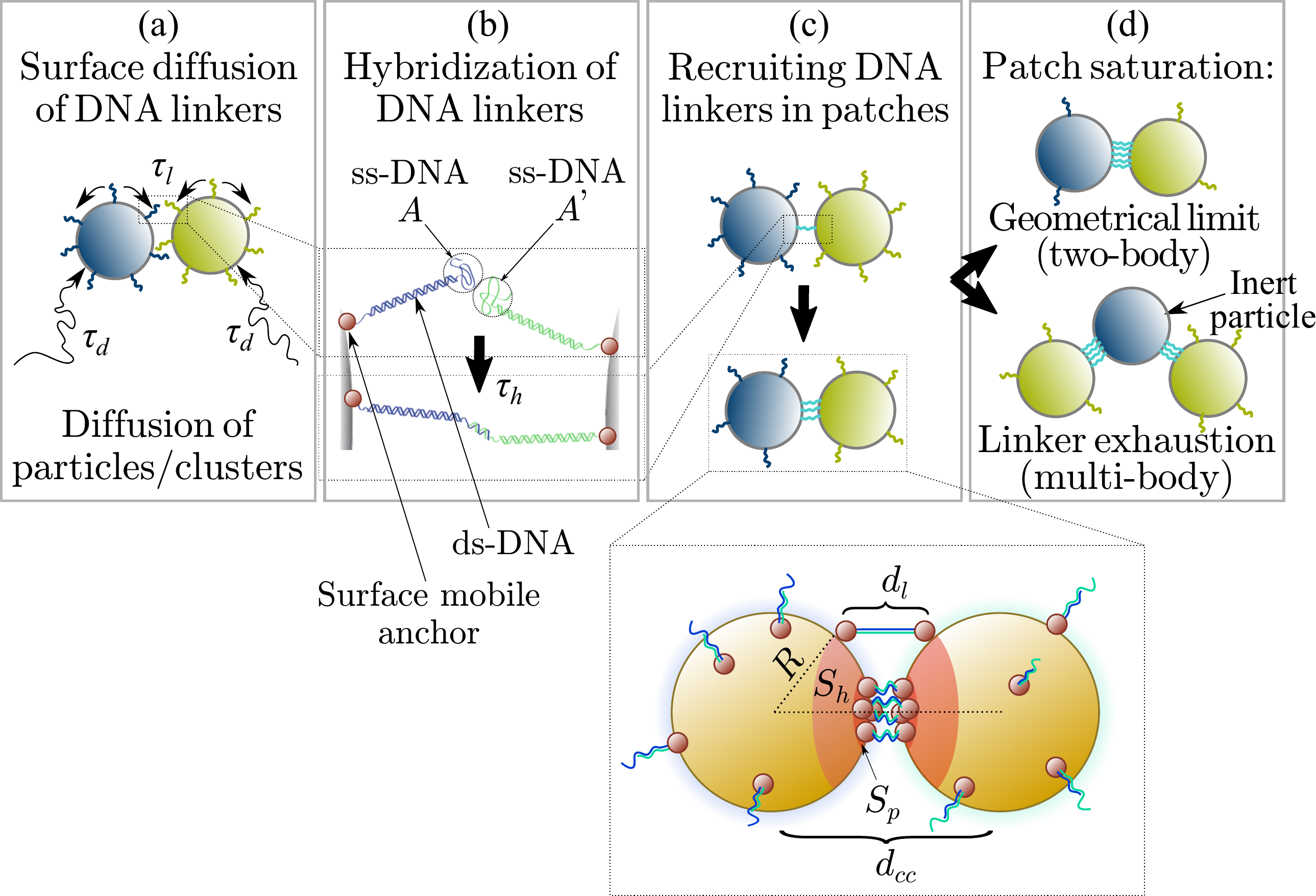}
\caption{\label{fig:model} Sketch of the main dynamic processes involved in the stable self-assembly of DNACCs by hybridization of surface mobile DNA linkers. (a) Hybridization is allowed by diffusion of the core particles and the surface diffusion of their linkers. (b) Structure of complementary DNA mobile linkers. (c) Recruiting of free linkers into bonding patches and detail of the geometry of a pair of crosslinked particles (lower sketch). In the latter, dark areas correspond to the `hybridization region', $S_h$, and the bonding patch, $S_p$. (d) Possible saturation mechanisms of the recruiting processes.}
\end{figure}

Fig.~\ref{fig:model} sketches the basic structure of DNACCs with surface mobile linkers and the main dynamic processes involved in their self-assembly. Typically, surface mobile DNA linkers are hybrid molecules consisting of a single strand DNA segment (ss-DNA) with prescribed sequence, a double strand DNA segment (ds-DNA) that acts as a semiflexible stem of the former and an anchor molecule adsorbed on the surface of the core particle (see Fig.~\ref{fig:model}b). The simplest self-assembling system one can make out of core particles functionalized with such mobile linkers is a binary symmetric suspension in which half of the particles carry linkers with a given ss-DNA sequence, $A$, and the other half carry the corresponding complementary sequence, $A'$, being otherwise fully equivalent. Without loss of generality, here we will use our model to address such simple symmetric $AA'$ self-assembling system, following the experiments of reference \cite{2018-mcmullen-prl}. In addition, we will focus on stable self-assembly conditions only, \textit{i.e.}, at temperatures well below the threshold hybridization temperature, $T \ll T_h$.

For an initially homogeneous system, diffusion will bring pairs of $AA'$ particles into a close range. This will allow their complementary linkers to meet and eventually hybridize while diffusing within a limited interstitial region of the particles surfaces (see Fig.~\ref{fig:model}a to \ref{fig:model}c). The maximum size of such `hybridization region' (dark areas in the detailed sketch of the particle pair in Fig.~\ref{fig:model}c) can be estimated for spherical core particles by simple geometrical considerations as a function of the radius of the core, $R$, the contour length of the linkers, $d_l$, and the interparticle center-to-center distance, $d_{cc}$. If we disregard any deformation of the core particles due to their bonding interactions \cite{2013-feng-sm}, which one can assume to be valid at least for solid cores and/or low number of crosslinkers, then the hybridization region is bounded by a spherical cap with surface area
\begin{equation}
S^0_h \approx \pi R (2R + d_l - d_{cc}) . %= \pi R (d_l - a),
\label{eq:S0h}
\end{equation}
Interestingly, very recent experiments have shown the validity of neglecting deformations even for soft core particles \cite{2021-mcmullen-pnas}. After the first stable hybridization takes place, the particles can be assumed to remain permanently bonded. However, further crosslinkers between them can be established as long as it is still possible to find pairs of free linkers diffusing within the hybridization region. This process, usually addressed as linker `recruiting', is limited by two conditions, outlined in Fig.~\ref{fig:model}d. First, one can assume that configurational entropy favors the grouping of the crosslinkers into relatively compact `bonding patches' occupying the central part of the hybridization region. As a consequence, the effective area of such region, $S_h$, decreases as the area of the enclosed bonding patch, $S_p$, grows, fulfilling at any time $S_h + S_p = S^0_h$. Thus, the size of any patch should saturate as it is geometrically limited  by the condition $S_p \le S^0_h$. Second, any patch growth on a given particle will also stop after all its free linkers have been recruited. At that point, the particle becomes `inert', as it is not able to bind to new additional neighbors anymore. In difference with the geometrical limit for the recruiting process of a given patch, which only depends on the properties and configuration of the concerned pair of particles, the exhaustion of free linkers on a particle with several patches depends on the competition between the recruiting processes of the latter. Therefore, if the concentration of linkers is low enough to make exhaustion more important than geometrical saturation, the effective valence of the particles is determined by many-body effects. Importantly, in difference with systems of rigidly anchored linkers \cite{2014-angioletti-uberti-prl, 2015-lequieu-sm}, many-body effects here depend not only on the particle configurations but also strongly on their history.

\section{\label{sec:model}Dynamic model}
In order to set up a model not only accurate but also computationally efficient, first one needs to identify the representative time scales of each elementary process among the ones outlined above. These are the diffusion of the particles in the background fluid, with characteristic time scale $\tau_d$, the diffusion of the linkers on their surfaces, with corresponding time scale $\tau_l$, and the characteristic time it takes the ss-DNA segments in close range to hybridize, $\tau_h$. In a typical system, initially one can expect $\tau_d \gtrsim \tau_l > \tau_h$ whereas at long times, when most particles are already assembled into more slowly diffusing clusters, $\tau_d \gg \tau_l > \tau_h$. Thus, in order to be able to approach the long self-assembly time scales observed in experiments \cite{2013-feng-sm, 2018-mcmullen-prl}, we adopt the strategy of treating explicitly the processes with slowest dynamics only, \textit{i.e.}. the diffusion of particles and clusters. However, we treat linker diffusion and hybridization processes implicitly, as probabilistic events ruled by effective dynamics.

For the explicit part of the model, we stick to the simplest representation of the particles---spheres of radius $R$---with only two explicit pair interactions: a repulsive soft core potential \cite{1971-weeks} and an elastic bonding potential between crosslinked neighbors \cite{1986-grest-pra}. These particles are labelled either as $A$ or $A'$, according to the type of implicit linkers they are associated to, and move under the action of Brownian forces corresponding to the thermal energy of the background fluid and the viscous friction, so that their diffusion constant is $D = kT /(6 \pi \eta R)$, where $k$ is the Boltzmann constant, $T$ the system temperature and $\eta$ the dynamic viscosity of the fluid. The corresponding dynamics is integrated by means of a Langevin dynamics (LD) scheme \cite{1986-grest-pra, 1987-allen}. Technical details of this scheme are provided in the \atsupp.

The implicit part of the model is taken into account after each LD integration step and includes three main aspects. First, we search for all non-crosslinked $AA'$ particle pairs whose center-to-center distance is smaller than $d_{cc} < 2R + d_l$, being thus candidates to establish their first crosslink. It is reasonable to assume that the probability of establishing a first crosslink decreases with $\tau_h$ and grows with the number of potential pairs of complementary linkers inside the hybridization region. The latter can be estimated by considering that diffusion tends to distribute homogeneously the available free linkers on a particle, $n_l$, over its available free surface, $S_{\mathrm{free}}$, which is simply the total core surface minus the surface occupied by all existing patches on the particle. Regarding the latter, reference \cite{2021-mcmullen-pnas} provides an accurate relationship between the number of crosslinkers in a patch and its characteristic size. Additionally, one has to keep in mind the time dependence of the parameters and the individual history of each particle in the pair. Thus, an accurate expression for the time probability of the first crosslink would require a rather complex analysis beyond the existing approaches for systems with rigidly anchored linkers \cite{2016-angioletti-uberti-pccp, 2019-jana-pre}. Here, however, we take a simple expression as a rough estimation of such probability. For a pair of particles 1 and 2, this is:
\begin{eqnarray}
P_{1, 2}(t) \approx && \, \min \left ( n_{l, 1}(t) \frac{S_h^0}{S_{\mathrm{free}, 1}(t)},\, n_{l, 2}(t) \frac{S_h^0}{S_{\mathrm{free}, 2}(t)}   \right ) \nonumber\\
&& \times \left ( 1 - e^{-\delta t / \tau_h}\right ),
\label{eq:firstlink}
\end{eqnarray}
where $\delta t$ is the LD integration time step. The first term implies that the probability is limited by the particle with lower number of linkers inside the hybridization region, whereas the second term places $\delta t$ as an attempting rate for a probabilistic event with characteristic time $\tau_h$. Here we can also take advantage of focusing on systems with small number of linkers only to approximate $S_{\mathrm{free}}$ as the total surface area of the particles at any time, $S_{\mathrm{free}, i}(t) \sim 4 \pi R^2$.

The second aspect addressed implicitly in our model is the growth of the bonding patches, which we represent by means of a dynamic function. Our strong assumption here is that this process is mainly determined by the surface diffusion of the linkers, with given diffusion constant $D_l$. The number of crosslinkers, $n_c$, in a patch established at time $t_0$ is expected to follow statistically a dynamic function that should grow at a characteristic rate, $\tau_l$, determined by $D_l$, and saturate to a maximum value, $n_c^{\mathrm{sat}}$, given by the aforementioned limits for the patch size. Besides these properties, we do not expect the qualitative behavior of the system to depend significantly on the specific form of such function. Therefore, we choose a simple expression for the patch growth rate, $d n_c/ dt = \tau_l^{-1} \left [ n_c^{\mathrm{sat}} - s(t) \right ]$, which gives
\begin{equation}
 n_c(t) = n_c^{\mathrm{sat}}(t) + \left [ n_c(t_0) - n_c^{\mathrm{sat}}(t) \right ] e^{-t/\tau_l}.
\end{equation}
For $\tau_l$ we take the empirical formula for the characteristic time of a particle diffusing on a sphere to reach an ideal circular adsorbing trap \cite{1981-chao-bpj},
\begin{equation}
 \tau_l \approx \log_{10} \left [ \frac{1}{1.55} \left ( \frac{2R + d_l - d_{cc}}{2R} \right )^{-2.26}\right ] \frac{R^2}{D_l},
\end{equation}
and for $n_c^{\mathrm{sat}}(t)$ we take
\begin{equation}
 n_c^{\mathrm{sat}}(t) = \min\left ( n_{l,\, 1}(t),\, n_{l,\, 2}(t),\, n_c^{\mathrm{max}}\right ),
 \label{eq:ncsat}
\end{equation}
where $n_c^{\mathrm{max}}$ is the number of crosslinkers corresponding to the geometrical limit of the patch size, $S_h^0$. As already underlined, here we will focus on the case $n_{l,\, 1}(t),\, n_{l,\, 2}(t) <  n_c^{\mathrm{max}}$ only.

\section{\label{sec:results}Results and discussion}
As reference for the validation of our model, we will compare our simulation results to the experimental measurements of McMullen and co-workers under kinetically controlled self-assembly conditions \cite{2018-mcmullen-prl}. Thus, the system we simulate consists in a monolayer of a symmetric binary mixture of microparticles with moderate area concentration and low surface densities of mobile linkers, whose hybridization threshold is well above the background (room) temperature. Table~\ref{tab:params} of the \atsupp\ includes all the relevant physical parameters of this system, directly taken from reference \cite{2018-mcmullen-prl} when available or from other similar experimental works otherwise.

\begin{figure}[t]
\includegraphics[width=0.95\columnwidth]{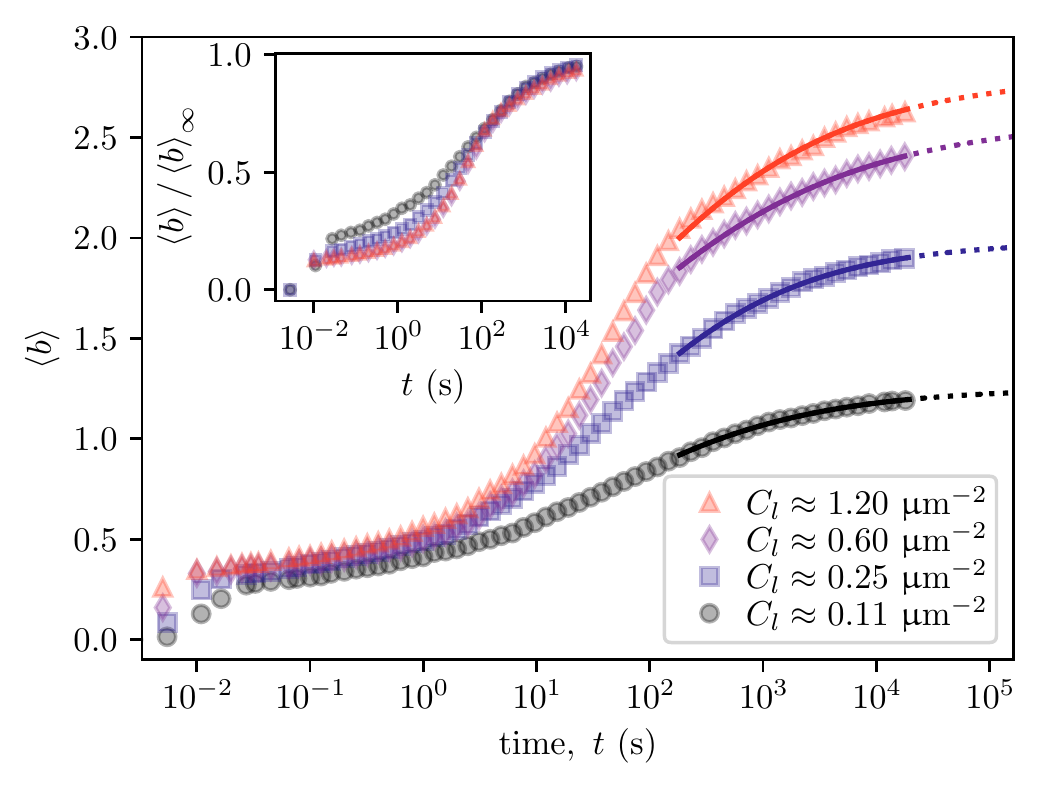}% Here is how to import EPS art
\caption{\label{fig:b-vs-time-Cl}Average coordination number of the simulated monolayers, $\langle b \rangle$, as a function of time for area concentration of particles $\sigma_d = 0.2$, $\tau_h = 0.02$~s and different number surface densities of linkers, $C_l$. Symbols correspond to the simulation data, solid/dotted lines to the fitted function \ref{eq:logistic}. Fitting is performed using the data corresponding to the solid part of the lines only. Inset shows the rescaling of the simulation curves with their fitted asymptotic value, $\langle b \rangle_{\infty}$.}
\end{figure}
In order to perform the aforementioned comparison, we sample different values of number surface densities of linkers, $C_l$, and characterize the self-assembly process by means of the effective valence of the particles, measured as their average coordination number, $\langle b \rangle$, in the long time limit, $\langle b \rangle_{\infty}$. Note that $C_l$ was not directly measured in the reference experiments, but only the buffer concentration of DNA during the preparation of the particles, $C_{\mathrm{DNA}}$, assuming a direct relationship between the latter and the final number of adsorbed linkers. Therefore, in absence of a defined correspondence between $C_l$ and $C_{\mathrm{DNA}}$, we first performed several simulation trials to find a range of values of $C_l$ that could reproduce the observed experimental structures---dimers, chains and networks, as shown in Fig.~2 of reference~\cite{2018-mcmullen-prl}---assuming a correct matching for the rest of parameters. The simulation results obtained with the selected set of linker densities are shown in Fig.~\ref{fig:b-vs-time-Cl} in logarithmic time scale. As one can see, after a short initial transient all curves display a sigmoid-like shape, with an initially slow growth weakly dependent on the linker concentration, followed by an inflection point around $t \sim 20$~s and a final saturation with a  strong dependence on $C_l$. This sigmoid-like shape, characteristic of systems with a limited growth behavior, implies that to further approach the asymptotic limit after the inflection point gets increasingly costly, becoming prohibitive for any direct dynamic simulation approach when the system gets very close to saturation. Therefore, the simulations were run up to $\sim 2\cdot10^4$~s only, time at which closeness to saturation is clear for all cases. In order to obtain a proper estimation of the expected saturation values of these curves, $\langle b\rangle_{\infty}$, we fitted the following logistic-like function,
\begin{equation}
\langle b \rangle (t \rightarrow \infty) = \langle b \rangle_{\infty}\left ( 1 - \frac{1}{1 + \left ( t / t_0 \right )^p}\right ),
 \label{eq:logistic}
\end{equation}
to the simulation data at times $t > 180$~s, consistently obtaining values for the least relevant fitting parameters around $t_0 \sim 20$~s---which would correspond to the inflection point of the curves---and $p \sim 0.4$. Regarding the most important parameter, the inset in Fig.~\ref{fig:b-vs-time-Cl} shows that the fitted values of $\langle b\rangle_{\infty}$ provide an excelent asymptotic rescaling of the curves, evidencing that the long time dynamics of the system depends quantitatively but not qualitatively on the linker density.
\begin{figure}[t]
\includegraphics[width=0.95\columnwidth]{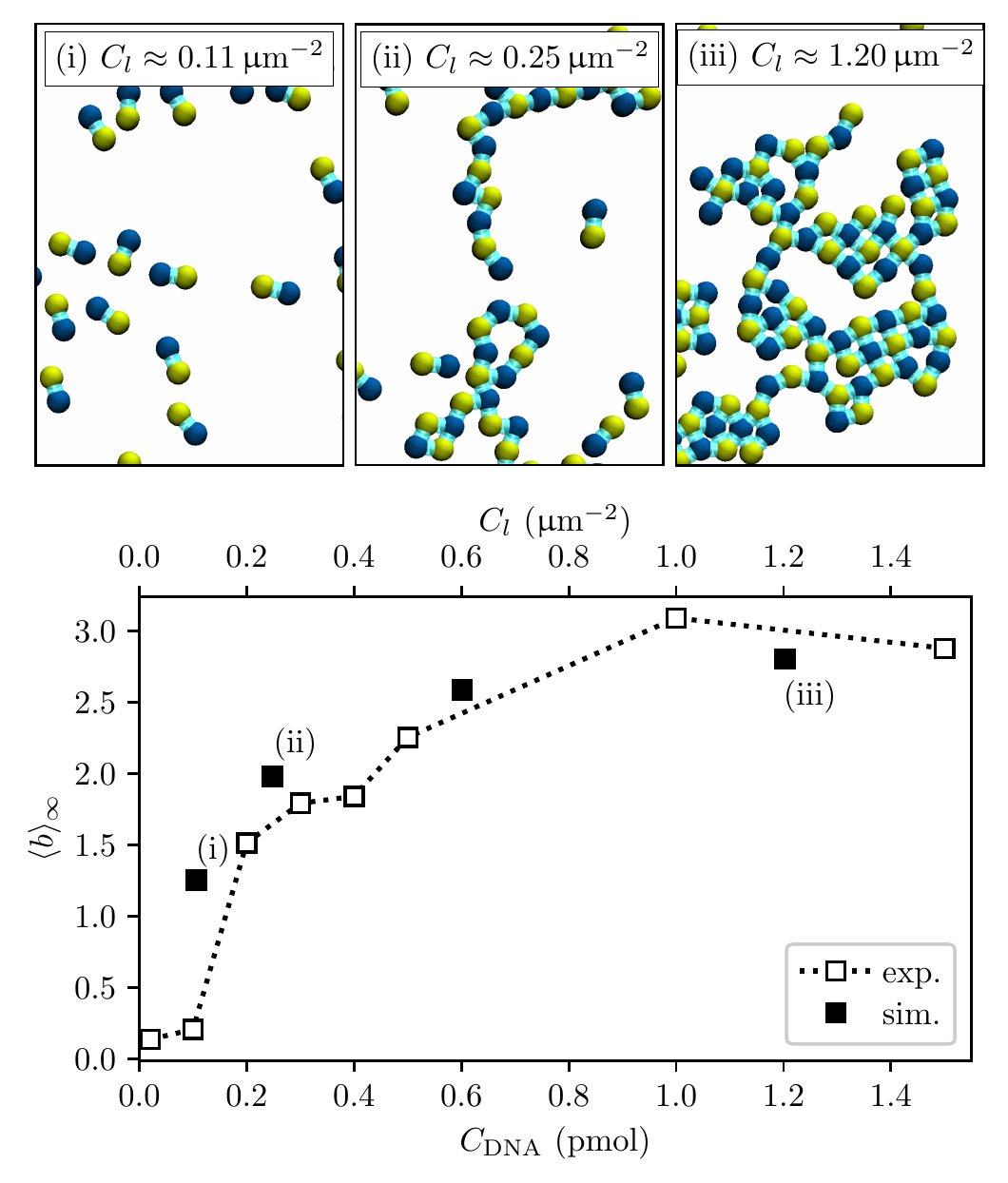}
\caption{\label{fig:exp-sim}Upper row: long time simulation snapshots obtained for selected linker densities. Lower panel: direct comparison of the average coordination number, $\langle b \rangle_{\infty}$, obtained in simulations (filled symbols) and corresponding experimental measurements from reference \cite{2018-mcmullen-prl} (empty symbols). Dotted line is a guide for the eye.}
\end{figure}

At this point, we can validate our simulation results by performing a direct comparison with experimental data. The upper row of Fig.~\ref{fig:exp-sim} shows several final simulation snapshots, corresponding to different values of $C_l$, that qualitatively match very well the experimental snapshots in Fig.~2 of reference \cite{2018-mcmullen-prl}. A more formal comparison is done in the lower panel, where the values of $\langle b \rangle_{\infty}$ calculated from simulations are shown together with the experimental ones, assuming a direct correspondence between $C_l$ (measured in $\upmu$m$^{-2}$) and $C_{\mathrm{DNA}}$ (measured in pmol). Despite the simplicity of this assumption, the agreement of the results is not only qualitatively but also quantitatively rather good, except for a significant overestimation in the value provided by the simulations for the smallest sampled linker concentration.

\begin{figure}[t]
\includegraphics[width=0.95\columnwidth]{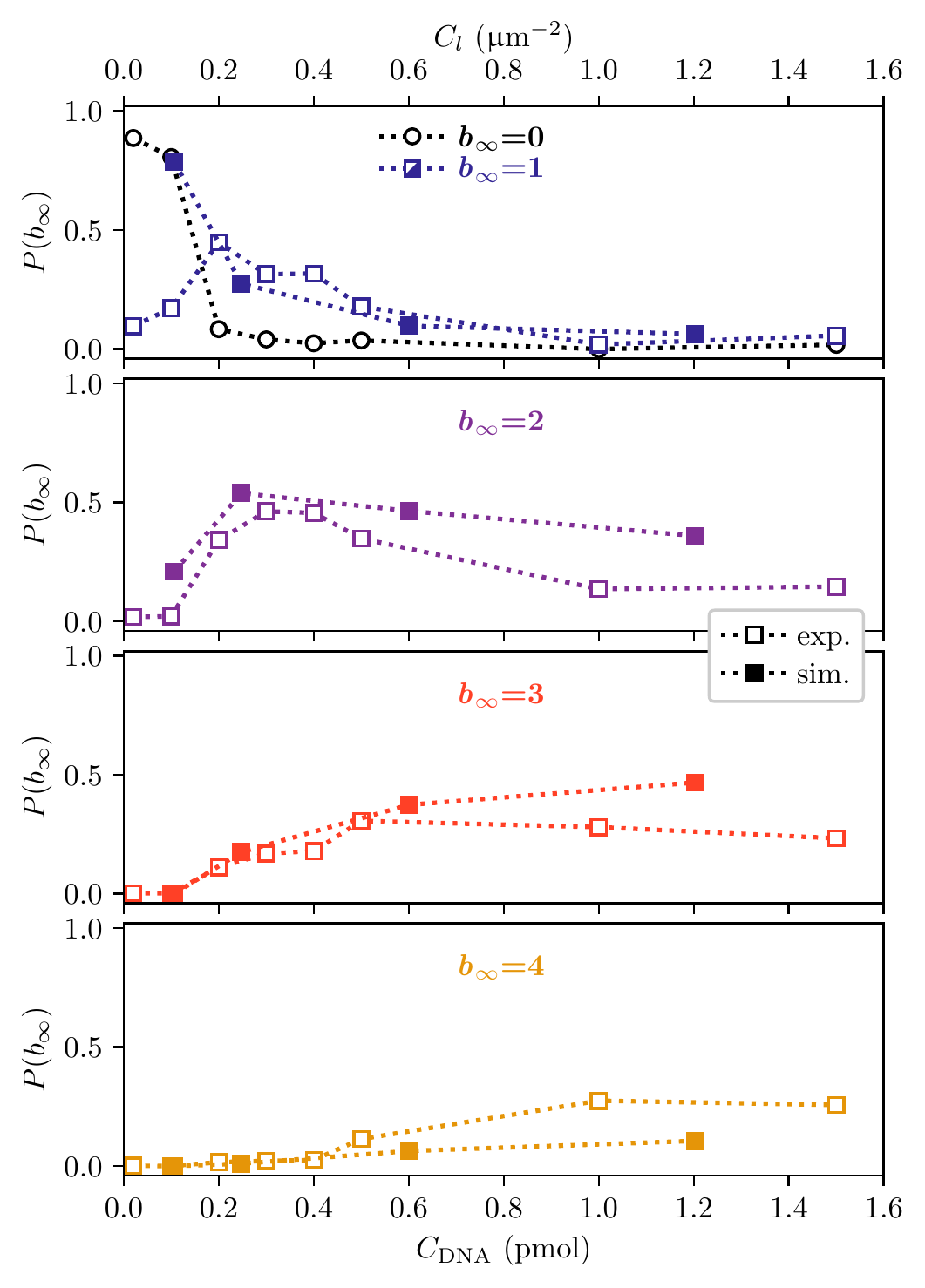}
\caption{\label{fig:exp-sim-probs}From top to bottom: split simulation results for the probability distributions of given values of the asymptotic coordination number, $P(b_{\infty} = \lbrace 1,\, 2, \, 3,\, 4\rbrace)$, as a function of the linker concentration (filled symbols), compared to the experimental measurements from reference \cite{2018-mcmullen-prl} (empty symbols). Top panel also includes the experimental distribution $P(b_\infty = 0)$ (empty circles). Probability of $b_{\infty}=0$ in simulations is negligible for any sampled $C_l$. Dotted lines are a guide for the eye.}
\end{figure}
An even more detailed comparison with experiments can be performed by calculating the probability distribution of a given asymptotic coordination number, $b_{\infty}$, as a function of the linker density. Fig.~\ref{fig:exp-sim-probs} shows the outcome of this comparison for coordination numbers in the range $b_{\infty} \in [0, 4]$. In general, there is a reasonably good quantitative agreement between simulation and experimental results, particularly at intermediate linker concentrations. The main discrepancy is the case $b_{\infty}=0$: whereas in experiments one can observe a large probability of finding non-crosslinked particles at very low values of $C_{\mathrm{DNA}}$, the simulation model provides no reason for such particles to not tend to vanish at long enough times for any finite $C_l > 0$. Density fluctuations may lead to some degree of asymmetry in the self-assembly process of a binary symmetric mixture, leaving a fraction of particles of one type unpaired in the long run, but it is very unlikely that such fraction could be so high as the one observed in experiments only for this reason. A more reasonble but still simple explanation for the discrepancy between experimets and simulations at low linker concentrations could be the existence of some degree of variation in the density of linkers of the experimental particles. For systems prepared under very low buffer concentrations, this would leave a significant fraction of particles without adsorbed linkers, thus being unable to experience any self-assembly. In addition, the correspondence between $C_{\mathrm{DNA}}$ and $C_l$ could be not exactly linear within the sampled range. The latter could also be the reason for the moderately bigger differences between simulation and experimental distributions that one can see at high linker densities. Nevertheless, the overall good agreement between experimental and simulation data is a strong indication of the validity of the model, which seems to capture very well the fundamental interplay between the main dynamic processes governing the self-assembly.

Once we proved the validty of our model, it is worth to use our results to analyze in more detail the system dynamics. As outlined above, the limited growth curves that characterize the self-assembly, introduced in Fig.~\ref{fig:b-vs-time-Cl}, indicate the existence of two main dynamic regimes separated by an inflection point. At early times, the growth of the coordination number following a regime that is both, slow and almost independent from the linker concentration, suggests that the self-assembly is dominated by the rate of `collisions'---\textit{i.e.}, approachings within the hybridization range---between $A$ and $A'$ non-crosslinked particles, or singlets. This would explain why the early behavior for the lowest linker concentration, that basically leads to the formation of dimers only, is so similar to the one corresponding to higher linker concentrations, for which larger clusters are possible. Note that the collision rate between singlets at early times is determined by the concentration of particles and their diffusion rate. However, the effects of the linker concentration, which in this stage enter the interactions through the early values of $n_{l,i}$ in the first term of Eq.~\ref{eq:firstlink}, turned out to be much less important. On the other hand, at long times the self-assembly behavior has to be dominated by the onset of the saturation mechanism, in this case the exhaustion of free linkers. Naively, one could expect the approach to saturation to happen at later times for those systems with higher linker concentrations, as the latter favor the formation of larger clusters with slower diffusion and, presumably, with a lower rate of new collisions that could increase the coordination. However, high coordinations enhance multi-body effects, increasing the rate of linker exhaustion and thus compensating the initially higher number of free linkers and the slowing down of the diffusion dynamics.
\begin{figure}[t]
\includegraphics[width=0.95\columnwidth]{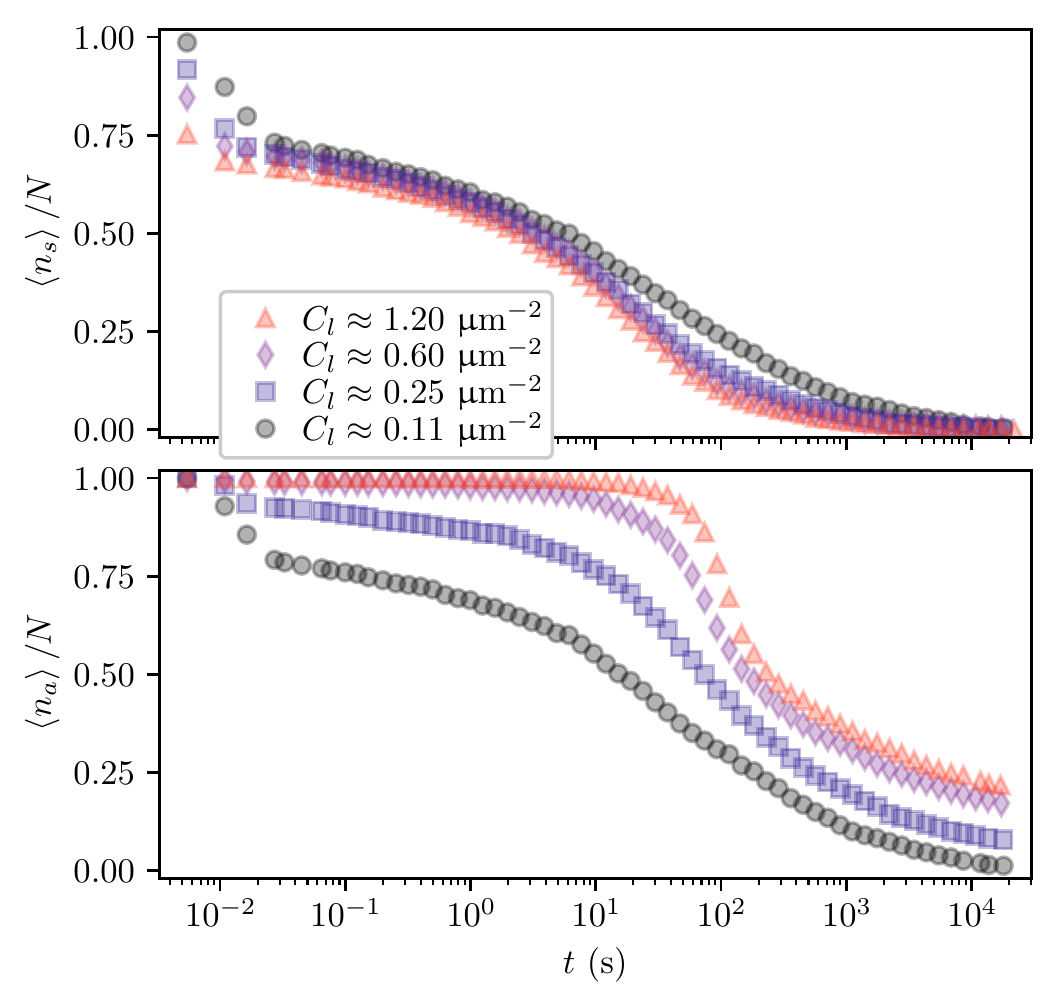}% Here is how to import EPS art
\caption{\label{fig:singact}Time evolution of the average fraction of non-crosslinked particles, or singlets, $\langle n_s \rangle/N$ (upper panel), and average fraction of non-inert particles, $\langle n_a \rangle/N$ (lower panel), remaining in the system and corresponding to each sampled linker density, $C_l$.}
\end{figure}

The interpretation of the early and late time dynamics discussed above can be better supported by computing sensible parameters for each sampled linker concentration. Fig.~\ref{fig:singact} shows the time evolution of the average fraction of singlets, $\langle n_s \rangle / N$, and the average fraction of `active'---\textit{i.e.}, non-inert---particles, $\langle n_a \rangle / N$, where $N$ is the total number of particles in the system. For the first parameter, we can also identify the early and late dynamic regimes in its time evolution: initially, the fraction of singlets decreases slowly and with a very weak dependence on the concentration of linkers, it drops to small values at times around $t \sim 20$~s and finally becomes nearly zero at long times. However, $\langle n_a \rangle / N$ shows a significant dependence on $C_l$ at early and intermediate times. The latter correspond to the region of strong drops, which become narrower and shifted towards later times as $C_l$ increases. Finally, the derivatives of these curves (not shown) indicate that at long times the rate of decrease of $\langle n_a \rangle / N$ becomes very similar for all $C_l$. These observations confirm the main influence of singlets dynamics at short times and the compensation of the influence of $C_l$ by multi-body effects at long times.

\begin{figure}[!t]
\includegraphics[width=0.99\columnwidth]{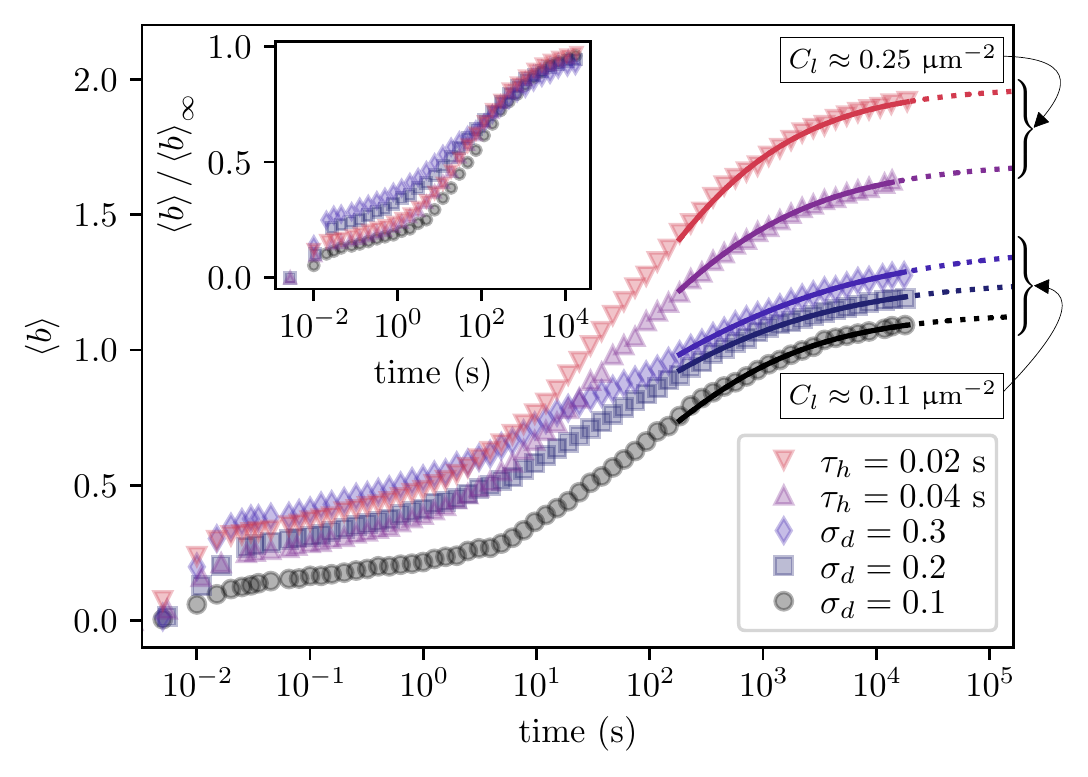}% Here is how to import EPS art
\caption{\label{fig:avb-time}Illustration of the influence of the hybridiation time, $\tau_h$, and particle area concentration, $\sigma_d$, on the average coordination number $\langle b \rangle$. The sampled values of $\tau_h$ and $\sigma_d$ correspond, respectively, to $\sigma_d=0.2$ and $\tau_h = 0.02$~s. Inset shows the rescaling of all the curves with their asymptotic value, $\langle b \rangle_{\infty}$, obtained from the same fitting procedure as in Fig.~\ref{fig:b-vs-time-Cl}.}
\end{figure}
Finally, we can take a step beyond the available experimental results and try to understand qualitatively the effects of other main parameters in the system. Specifically, here we briefly discuss the influence of the particle area concentration, $\sigma_d$, and the DNA hybridization time, $\tau_h$. Fig.~\ref{fig:avb-time} shows the time evolution of the average coordination number obtained for several systems with different values of $\sigma_d$ and $\tau_h$. To ease their observation, we split the curves by showing these parameters for two linker concentrations. First, we can observe that an increase in the hybridization time clearly leads to a decrease of the asymptotic coordination: by increasing $\tau_h$ from 0.02 to 0.04~s we obtain a decrease from $\langle b \rangle_\infty \sim 2.0$ to $\langle b \rangle_\infty \sim 1.7$ at $C_l \approx 0.25~\upmu\mathrm{m}^{-2}$. This is reasonable, as the longer average time it takes to form the first crosslink between a pair of colliding particles increases the chance of exhaustion of free linkers before the new crosslink is created if any of the involved particles has already some recruiting patch with an existing neighbor. Regarding the particle area concentration, it is also clear its opposite effect on the asymptotic coordination: an increase of $\sigma_d$ leads to an increase of $\langle b \rangle_{\infty}$. This is a consequence of the decrease of the characteristic interval between favorable collisions led by the lower average distance between particles, which increases the rate of formation of dimers at early times and decreases the influence of multi-body effects at late times. Interestingly, in all cases we keep obtaining a collapse of the curves at long times by normalizing with their asymptotic value.

We conclude the discussion of our results with an important consideration. The observed good agreement with experimental results is a strong indication of the correctness of the relative importance of each dynamic process considered in our model. However, this does not guarantee that our individual quantitative assumptions on the reference system are correct. For instance, we could have understimated significantly the effective value of the hybridization time, $\tau_h$, a parameter that is particularly difficult to estimate. This would led us to sample linker concentrations actually smaller than the experimental ones while keeping a correct ratio between such parameters. Thus, we evidenced that our model properly captures the competing interplay of the main dynamic processes that determine the self-assembly in these systems.

\section{\label{sec:conclusions}Conclusions and outlook}
We presented the first model for the stable self-assembly of suspensions of DNACCs with surface mobile linkers that properly captures the interplay between the main dynamic processes governing the system under kinetically limited valence conditions. This corresponds to linker concentrations low enough to make multi-body effects the saturation mechanism determining the asymptotic values of the average coordination number.

Our model has been validated by direct comparison with experimental results for a monolayer of a symmetric binary mixture of self-assembling particles, showing a rather good quantitative agreement. In addition, our results provide evidences of the existence of two dynamic regimes during the self-assembly process: at short times, the system is mainly governed by the diffusion dynamics of non-crosslinked particles and the formation of dimers. At long times, the average coordination follows a universal, limited growth curve qualitatively independent of the system parameters. Finally, we outlined the qualitative effects of another two parameters not explored in the reference experiments: the characteristic linker hybridization time, whose increment leads to a decrease of the effective average valence, and the particle concentration, which favors its increase.

The modeling approach introduced here opens up the possibility to explore in simulations a broad range of variations of these systems at a modest computing cost. For instance, an extension of the simulations can be easily performed to consider three-dimensional systems and multiple combinations of complementary linkers. Finally, our model can be also extended by introducing the simulation of thermodynamic equilibrium for linker binding/unbinding at temperatures around the hybridization transition.

\section*{\label{sec:appendix}Appendix: simulation details}
\setcounter{table}{0}
\renewcommand{\thetable}{A\arabic{table}}
\begin{table*}[!h]
\caption{\label{tab:params}%
Physical parameters of the simulated system.
}
\begin{tabular}{lccc}
\toprule
\textrm{Parameter}&
\textrm{Value (S.I. units)}&
\textrm{Reference} &
\textrm{Value (reduced units)}\\
%\colrule
\midrule
Background fluid dynamic viscosity, $\eta$ & $8.9 \cdot 10^{-2}$~$\textrm{kg}/(\textrm{m} \cdot \textrm{s})$ & (water at room $T$) & $1.069 \cdot 10^3$\\
Particle radius, $R$ & $1.5 \cdot 10^{-6}$~m & \cite{2018-mcmullen-prl} & 0.5\\
Particle core density, $\rho$ & 965~$\textrm{kg}/\textrm{m}^3$& \cite{2018-mcmullen-prl} (PDMS at room $T$) & 0.968\\
Particle diffusion constant, $D_d$ & $1.64 \cdot 10^{-13}$~\textrm{m}$^2$/\textrm{s}& (Stokes-Einstein equation) & $9.9 \cdot 10^{-5}$\\
Crosslinker contour length, $d_l$ & $4.4 \cdot 10^{-8}$~m & \cite{2018-mcmullen-prl} \textrm{(estimated)}& $1.47 \cdot10^{-2}$\\
Crosslinker close packing size, $a$ & $5 \cdot 10^{-9}$~m & \cite{2018-mcmullen-prl} \textrm{(estimated)}& $1.67 \cdot 10^{-3}$\\
Surface diffusion constant of linkers, $D_l$ & $2 \cdot 10^{-13}$~ \textrm{m}$^2$/s & \cite{2016-joshi-sa} & $1.21 \cdot 10 ^{-4}$\\
\bottomrule
\end{tabular}
\end{table*}
The diffusion of the DNACCs in the background fluid is calculated by means of the Langevin dynamics simulation approach, \textit{i.e.}, molecular dynamics simulations in the NVT ensemble with a Langevin thermostat. The latter provides an implicit representation of the thermal fluctuations of the background fluid \cite{1987-allen}. This approach requires smooth interaction potentials between the particles. Thus, we represent the excluded volume interactions for spherical DNACCs of radius $R$ as a Weeks-Chandler-Andersen (WCA) soft-core potential \cite{1971-weeks}:
\begin{equation}
U_{\mathrm{{WCA}}}(r)=\left\{ \begin{array}{ll}
U_{\mathrm{{LJ}}}(r) - U_{\mathrm{{LJ}}}(r_{\mathrm{cut}}), & r<r_{\mathrm{{cut}}}\\
0, & r\geq r_{\mathrm{{cut}}}
\end{array}\right. ,
\label{eq:WCA}
\end{equation}
where $U_{LJ}(r) = 4\epsilon_s \left[\left(2R/r\right)^{12} - \left(2R/r\right)^{6}\right]$ is the conventional Lennard-Jones potential and $r_{\mathrm{cut}} = 2^{7/6}R$. DNA crosslinks between particles are represented implicitly by a finitely extensible non-linear elastic (FENE) bonding potential \cite{1986-grest-pra},
\begin{equation}
U_{\mathrm{FENE}}(r) = - \frac{K_{b}}{2} r^2_{\mathrm{max}} \ln \left [ 1 - \left ( \frac{r}{r_{\mathrm{max}}}\right )^2 \right ],
\label{eq:FENE}
\end{equation}
with $r_{\mathrm{max}} = 3R$. For the choice of the prefactors in potentials (\ref{eq:WCA}) and (\ref{eq:FENE}), we assumed that it can be arbitrary as long as it provides small fluctuations of the center-to-center distance of the bonded particles around $2R$. Here, we chose $\epsilon_s = 10~kT$ and $K_b = 7.5~kT/R^2$. The motion of each particle $i$ in the system is calculated by integrating the Langevin equation of motion with a velocity-verlet scheme. The Langevin equation is obtained by adding stochastic and friction terms to the Newtonian equation of motion:
\begin{equation}
m_i \frac{d{\vec v}_i}{dt} = {\vec F}_i - \Gamma_\mathrm{T} {\vec v}_i + (2 \Gamma_T kT)^{1/2}{\hat \xi}_{i},
\label{eq:langevin}
\end{equation}
where $m$ and $\vec v_i$ are respectively the mass and velocity of the particle, $\vec F_i$ the net force acting on it, $\Gamma_T = 6 \pi \eta R$ its friction constant and ${\hat \xi}_{i}$ is a random unitary vector with uncorrelated components, thus providing a stochastic force that follows the usual fluctuation-dissipation rules \cite{1987-allen}.

Finally, Table~\ref{tab:params} summarizes the physical parameters of the systems under study. These consisted in monolayers of 256 particles of type $A$ and 256 of type $A'$ initially mixed in random non-overlapping configurations inside a quasi two-dimensional simulation box with lateral periodic boundaries. In order to ensure the numerical stability of the integration scheme, a system of reduced units was chosen by taking $kT = 1$, $m = 1$ and $R = 0.5$. Following this system, integration time step was set to $\delta t = 9 \cdot 10^{-5}$.

All the simulations were performed using the {ESPResSo} 4.1.4 simulation package \cite{2019-weik-epjst}. Results presented here correspond to averages over 5 independent runs.

\section*{\label{sec:ack}Acknowledgments}
Simulations were partially performed at the Vienna Scientific Cluster (VSC). SSK acknowledges partial support of RFBR-DFG grant N 21-52-12013.

\end{document}